# Element-Specific First Order Reversal Curves Measured by Magnetic Transmission X-ray Microscopy

Dustin A. Gilbert,[1] Mi-Young Im,[2] Kai Liu,[3] Peter Fischer[2,4]

[1]Materials Science Department, University of Tennessee, Knoxville, TN 37996, USA

[2]Materials Sciences Division, Lawrence Berkeley National Laboratory, Berkeley, CA 94720, USA

[3]Physics Department, Georgetown University, Washington DC 20057, USA

[4]Physics Department, University of California, Santa Cruz, CA 95616, USA

**Abstract**: The first order reversal curve (FORC) method is a macroscopic measurement technique which can be used to extract quantitative, microscopic properties of hysteretic systems. Using magnetic transmission X-ray microscopy (MTXM), local element-specific FORC measurements are performed on a 20 nm thick film of CoTb. The FORCs measured with microscopy reveal a step-by-step domain evolution under the magnetic field cycling protocol, and provide a direct visualization of the mechanistic interpretation of FORC diagrams. They are compared with magnetometry FORCs and show good quantitative agreement. Furthermore, the high spatial resolution and element-specific sensitivity of MTXM provide new capabilities to measure FORCs on small regions or specific phases within multicomponent systems, including buried layers in heterostructures. The ability to perform FORCs on very small features is demonstrated with the MTXM-FORC measurement of a rectangular microstructure with vortex-like Landau structures. This work demonstrates the confluence of two uniquely powerful techniques to achieve quantitative insight into nanoscale magnetic behavior.

**Introduction**

Research and design of modern magnetic materials frequently requires a microscopic understanding of the structure and properties of the system,[1-3] which can be challenging to determine with macroscale measurements.[4-6] On the other hand, advanced magnetic imaging, such as magnetic transmission X-ray microscopy (MTXM)[7-12], Lorentz transmission electron microscopy,[13] spin-polarized scanning tunneling microscopy,[14,15] spin-polarized low energy electron microscopy (SPLEEM),[16,17] photoemission electron microscopy,[18,19] and scanning electron microscopy with polarization analysis,[20-22] can provide critical visualization of real space magnetic configurations. However, magnetic imaging over local areas face limitations to effectively and quantitatively resolve the sometimes large variations in materials properties of a system. An ideal research technique would bridge these two domains: quantitative macroscale measurements with exemplary spatial resolution or microscopy that can quantitatively survey magnetic property variations.[6]

The first order reversal curve (FORC) technique has shown promise as a bridge of the former type.[9,23-28] This technique analyzes the evolution of the magnetization along a series of partial hysteresis loops – macroscopic measurements – to extract magnetic characteristics such as microscopic interaction fields and intrinsic coercivity distributions. These properties often are not associated with any particular regions of the sample, but represent sample-scale ensemble-averaged distributions. Furthermore, FORC distributions have been used extensively to provide insights into the magnetization reversal processes,[9,29-38] yet direct confirmation with magnetic imaging, particularly one carried out using the same magnetic field cycling protocol, has been lacking. In fact, performing FORC measurements with microscopy which is sensitive to magnetic structure provides a mechanism to extract quantitative insights from microscopic probes, becoming a bridge of the latter type.

In this work, we combine the FORC technique with MTXM to quantitatively evaluate magnetization reversal behavior.[26-28,39,40] Combining these two techniques allows the magnetic interaction fields and intrinsic coercivity distributions to be quantitatively extracted from local areas. MTXM further provides the ability to highlight specific regions or phases[3] using its element-specific, high-resolution[11] imaging capability. The MTXM-FORCs measured on a CoTb film with perpendicular anisotropy are directly compared to FORCs measured on the same sample with alternating gradient magnetometry (AGM). Since the FORC technique has been proposed as a macroscopic measurement which can "fingerprint" microscopic reversal behavior,[9,29-38,41] comparing FORC measurements captured with macroscopic techniques and microscopy presents an ideal opportunity to confirm this capability.[42,43] The results of this work lay the foundation for broader applications of FORC and MTXM, including applying the FORC technique to exceedingly small samples to quantitatively extract microscopic information, as demonstrated on a single permalloy (Py) microstructure with vortex-like flux-closure structures.

**Experiment**

In the first sample, a single magnetic film of Pt(5 nm)/CoTb(20 nm)/Pt(5 nm) with perpendicular magnetic anisotropy (PMA) was grown by sputtering on an X-ray transparent amorphous $Si_3N_4$ (200 nm) membrane in a 5 mTorr Ar atmosphere, in an ultrahigh vacuum sputtering chamber. The Pt layers were used as an adhesion/seed layer and capping layer, respectively. MTXM images were obtained in the out-of-plane geometry on the XM-1 microscope (BL 6.1.2) at the Advanced Light Source using 778 eV X-rays, probing the Co $L_3$ edge.[7] Fresnel zone-plate optics were used for X-ray focusing and imaging, and the full field images were recorded by an X-ray sensitive CCD camera. Magnetic contrast was obtained using the XMCD effect at the Co $L_3$ edge. The use of XMCD makes this technique inherently element-specific. For the present CoTb system, which consist of only a single magnetic phase, the element specific sensitivity does not provide an additional insight, but it can be used to separate the magnetic signals in a system with more than one magnetic material or phase. To obtain FORC data, XMCD images were recorded at each magnetic field step following the FORC sequence described below. From the images, the area of the black and white contrasted domains, $A_{White}$ and $A_{Black}$ were measured, corresponding to the positive and negative out-of-plane directions, respectively. A normalized magnetization is defined as $M_{Mic} = \frac{A_{White} - A_{Black}}{A_{White} + A_{Black}}$. Magnetometry measurements were performed at room temperature using an alternating gradient magnetometer (AGM). The magnetization from the AGM is identified by the variable $M$.

In the second sample, a single 40 nm thick, rectangular (3 μm × 2 μm) microstructure of permalloy ($Ni_{80}Fe_{20}$) with in-plane magnetization was prepared on a $Si_3N_4$ windows using electron-beam lithography and liftoff techniques. For this size and shape of microstructure, the magnetization curls into a pair of vortex structures, e.g. Landau patterns, with opposite circularities; circularity is defined as the direction of the in-plane winding of the chiral structure and can be clockwise or counter-clockwise.[22,44] MTXM images were obtained using circularly polarized 708 eV X-rays, probing the Fe $L_3$ edge.[8,34] Imaging was performed in a tilted geometry, with the X-rays impinging at 30° relative to the sample normal, capturing the in-plane component of the magnetization. The black and white contrasts indicate regions with magnetization parallel and antiparallel to the positive magnetic field direction, respectively. The magnetic sensitivity was enhanced by subtracting images taken with left- and right-circularly polarized X-rays. The magnetization from a single microstructure is calculated to be 180 pemu, which is beyond the sensitivity of conventional magnetometers.

First order reversal curve measurements for both MTXM and AGM were performed following a previously reported magnetic field sequence.[9,26,45] From positive saturation the applied magnetic field is reduced to a scheduled reversal field, $H_R$. At $H_R$ the field sweep direction is reversed and the magnetization, $M$ or $M_{Mic.}$, is measured as the applied field, $H$, is increased back to positive saturation. This process is repeated for $H_R$ between the positive and negative saturated states, thus measuring a family of FORCs.[9,31] A mixed second order derivative is applied to the dataset to extract the FORC distribution: $\rho(H, H_R) \equiv -\frac{1}{2}\frac{\partial}{\partial H_R}\left(\frac{\partial (M \text{ or } M_{Mic})}{\partial H}\right)$. Along each FORC branch (increasing $H$ at a constant $H_R$) the derivative $\frac{\partial (M \text{ or } M_{Mic})}{\partial H}$ captures domain growth or 'up-switching' events. The derivative $\frac{\partial}{\partial H_R}$ distinguishes new up-switching events on adjacent FORC branches.[26] It is thus important to note that the absence of any feature in the FORC diagram only indicates that the magnetization is changing at the same rate as that on adjacent FORCs (adjacent in $H_R$).

**Results**

*FORCs of a CoTb Film*

The families of FORCs and the corresponding FORC distributions for the CoTb sample are shown in Figure 1, measured by (a, b) AGM and (c, d) MTXM, respectively. FORC distributions from both techniques exhibit a boomerang feature, typical of PMA thin films that reverse by a domain nucleation and growth mechanism.[9,32] Along decreasing $H_R$, the horizontal ridge appears at $\mu_0 H_R$= -45 mT (highlighted as feature 1 in Fig. 1b) and corresponds to the initial nucleation and rapid propagation of reversed domains when the applied field is reduced from positive saturation.[8] As the field is increased from $H_R$ back to saturation, an elongated horizontal FORC feature emerges as the previously reversed domains return to positive saturation. The horizontal feature is accompanied, at more negative $H_R$, by a pair of vertical negative/positive FORC features (marked as features 2 and 3 in Fig. 1b), which have been associated with the irreversible domain annihilation process. Details of these features in the context of the MTXM images are explored below. The FORC diagram from the AGM is more finely defined, resulting from the many more FORC branches measured compared to the MTXM data.

Quantitative comparisons of the FORC features in Figs. 1b and 1d show reasonably good agreement. Specifically, the intensity of the FORC features 1-3, as determined by their maximum value, for the AGM FORC are 1 : 0.70 : 0.43, and for the MTXM FORC are 1 : 0.70 : 0.51, respectively. Furthermore, feature 1 is centered at ($\mu_0 H$=44 mT, $\mu_0 H_R$ =-45 mT) in the AGM FORC diagram (Fig. 1b) and at ($\mu_0 H$=44 mT, $\mu_0 H_R$ =-46 mT) in the MXTM FORC diagram (Fig. 1d), respectively, showing good agreement. Similar consistency is observed in the locations of features 2 and 3 as well. Together, these results confirm that FORCs measured by MTXM imaging, that probe the microscopic responses, are quantitatively comparable to FORC measurements performed by traditional magnetometry methods that capture the collective magnetic response from the entire sample. This agreement can be expected for measurements in-which the field of view (in this case the 10 μm diameter provided in the MTXM images) captures behavior that is representative of the system as a whole. This agreement illustrates that FORC measurements performed with conventional magnetometry also contain nanoscale details about the magnetization reversal.

To illustrate the step-by-step microscopic magnetization reversal process, Figure 2 shows the family of FORCs measured with MTXM (Fig. 2a), the AGM FORC diagram (Fig. 2b) and a selection of the MTXM images from the CoTb film (Fig. 2c). The AGM FORC diagram is used due to its higher resolution, while the MTXM family of FORCs is used because of its direct correspondence with the MTXM images, with each image in Fig. 2c marked by a star in Fig. 2a. Following the FORC measurement sequence, the film was first saturated, then the field was reduced to an $H_R$, and MTXM images were taken as the magnetic field was increased to saturation. Images taken on FORCs starting at $\mu_0 H_R$ > -40 mT show no magnetic contrast, indicating that that the magnetization remains in the saturated state at $H_R$; since $M$ is constant in the saturated region, the derivative $\frac{\partial M}{\partial H} = 0$ and is independent of $H_R$, resulting in no features in the FORC distribution. For this reason, the images taken at saturation are not shown. Reducing $\mu_0 H_R$ to -40 mT (top row of Fig. 2c), a single large black domain overtakes the lower ≈30% of the image, highlighted by the dashed yellow boundary; the black (white) features indicate a magnetic domain oriented into (out-of) the page. With increasing magnetic field (images from left to right in the figure), following the FORC measurement sequence, the black domain does not change appreciably between -40 mT and +20 mT, then shrinks slightly between 20 mT and 38 mT, and rapidly contracts between 38 mT and 55 mT. These magnetic behaviors appear for the first time along the FORC which starts at $\mu_0H_R$= -40 mT, just after the initial rapid propagation of reversal domains, and therefore manifest features in the FORC distribution specifically at $\mu_0H_R$= -40 mT. The slight creep (small $\frac{\partial M}{\partial H}$) followed by rapid contraction of the magnetic domain (large $\frac{\partial M}{\partial H}$) appears as a weak feature in the FORC diagram, starting at $\mu_0H$= 20 mT and achieving a maximum at $\mu_0H$= 44 mT, defining feature 1 (along the first horizontal dashed line in Fig. 2b). Further reducing $\mu_0H_R$ to -45 mT (second row in Fig. 2c), the black domain covers much more of the image, indicating more of the sample has reversed. Again, increasing $H$ along the FORC branch, the domain undergoes a relatively small contraction between $\mu_0H$= 20 mT and 38 mT, then a rapid contraction between 38 mT and 55 mT. Since the change in magnetization between $\mu_0H$= 20 mT and 55 mT is larger on the $\mu_0H_R$= -45 mT FORC compared to the $\mu_0H_R$=-40 mT FORC, the feature in the FORC distribution is again positive (along the 2$^{nd}$ horizontal dashed line in Fig. 2b). Next, starting from $\mu_0H_R$= -55 mT (middle row in Fig. 2c) the domain size at $H_R$ is similar to that in $\mu_0H_R$= -45 mT. Increasing $H$, the domain contraction follows a very similar progression. As a result, the FORC distribution exhibits a relatively weak feature (along the 3$^{rd}$ horizontal dashed line in Fig. 2b). These images offer a visual confirmation of the $\partial H_R$ derivative in the FORC diagram capturing the new domain growth events that are unique to a particular FORC branch. These results also confirm the conjecture of Davies et al.,[9] that the horizontal FORC feature (labeled 1 in Fig. 2b) identifies a rapid domain propagation at $H_R$, followed by steady domain evolution as $H$ is increased, and that the peak between $\mu_0H$=38 mT and 55 mT indicates a rapid growth of the positively oriented domains.

Next, the FORC branch starting at $\mu_0H_R$= -72 mT is considered (fourth row in Fig. 2c). At $H$=$H_R$, the sample exhibits only two small positively oriented domains (white), indicated by the yellow arrows – the field of view is mostly negatively saturated. Increasing from $\mu_0H$= -72 mT to 38 mT the domains expand slightly, qualitatively similar to the domains along the FORC which started at $\mu_0H_R$= -55 mT. Due to the fewer number of domains and their smaller size in the $\mu_0H_R$= -72 mT images, this corresponds to a smaller change in the magnetization, meaning $\frac{\partial M}{\partial H}$ is smaller along the $\mu_0H_R$= -72 mT FORC branch. Mathematically this *decrease* in $\frac{\partial M}{\partial H}$ in FORCs with successive $H_R$ results in a negative feature in the FORC distribution, labeled 2 in Fig. 2b.[26] From $\mu_0H$= 38 mT to $\mu_0H$= 55 mT the positively oriented domains (white) in the $\mu_0H_R$= -72 mT FORC branch expand significantly, becoming comparable in relative size to the $\mu_0H_R$= -55 mT FORC. The increase in the white domain area between 38 mT and 55 mT is larger along the $\mu_0H_R$= -72 mT FORC branch compared to the $\mu_0H_R$= -55 mT branch, corresponding to an *increase* in $\frac{\partial M}{\partial H}$ in successive $H_R$ and a positive feature in the FORC diagram. At subsequent fields along the $\mu_0H_R$= -72 mT FORC, the domain evolution – specifically the relative areas of the white and black domains – tracks closely with the $\mu_0H_R$= -55 mT, resulting in similar $\frac{\partial M}{\partial H}$ along the adjacent FORC branches. The lack of appreciable change in $\frac{\partial M}{\partial H}$ is manifested as no feature in the FORC distribution. Again, it is interesting to note that the local domain structure is very different, but the FORC distribution plots the ensemble average, in this case the field of view of 80 μm².

These same trends are observable in the FORC branch measured from $\mu_0 H_R$= -100 mT (bottom row in Fig. 2c), which has achieved an apparent negative saturation at $H=H_R$. As the sample remains largely negatively saturated between $\mu_0 H$= -100 mT and 38 mT, $\frac{\partial M}{\partial H}$ is essentially zero. Thus, the derivative along the $\mu_0 H_R$= -100 mT FORC branch is smaller than that along the $\mu_0 H_R$= -72 mT branch, further contributing to the negative FORC feature (bottom dashed line in Fig. 2b). Only during $\mu_0 H$=38 mT to 55 mT does the sample undergo reversal by domain nucleation and rapid propagation, generating a large $\frac{\partial M}{\partial H}$ compared to the $\mu_0 H_R$= -72 mT FORC branch, which is reflected in the positive FORC feature.

Thus the vertical FORC features (labeled 2 and 3 in Fig. 2b) are directly confirmed to correspond to large regions of the sample becoming negatively saturated and re-nucleating domains along the FORC branch. Achieving negative saturation suppresses domain growth between $H=H_R$ and the rapid propagation event at $\mu_0 H$=38 mT, resulting in the negative FORC feature (labeled 2 in Fig. 2b). Between $\mu_0 H$=38 mT and 55 mT domain nucleation and rapid propagation occurs, facilitating the positive FORC feature (labeled 3). At $\mu_0 H$>55 mT the domain progression is largely independent of $\mu_0 H_R$ and thus does not result in any appreciable feature in the FORC distribution.

Performing FORCs using MTXM provides a unique opportunity to capture the microscopic magnetic configurations, which are inaccessible with conventional magnetometry. These configurations are important because FORC captures the magnetic evolution from each reversal field, which are presumed to be evolved sequentially between neighboring $H_R$. Without this sequential evolution, the initial configuration can contribute to the shape of each FORC branch, presenting a new variable in deciphering the FORC distribution. This Ansatz is tested by comparing the magnetic configuration at each $H_R$ and subsequent evolution. Specifically, at $\mu_0 H_R$= -40 mT there is a large, single negative domain (outlined in yellow dashes in Fig. 2c); at $\mu_0 H_R$= -45 mT this domain is partly re-oriented back to positive (orange dashes); at $\mu_0 H_R$= -50 mT some of these regions are reoriented back to negative (blue dashes), meanwhile other regions which were negative at $\mu_0 H_R$= -45 mT have now become positive (orange dashes). This flip-flopping of the magnetization against the applied field is unlikely to be the result of sequential evolution but rather a consequence of random thermal activation at the nucleation field and transient domain evolution. The resulting domain configurations are local energy minima that the FORC protocol is able to access as the energy landscape is changed during the field cycling.[46] The domain evolution from each of these states is also shown to be different (Fig. 2c), confirming the Ansatz. However, the FORC diagrams for the conventional (whole sample) and microscopic FORCs were already shown to be very comparable, implying that the ensemble average accurately represents behavior down-to the scale of the MTXM field of view (80 $\mu m^2$), seemingly indifferent to the initial configuration at $H_R$.

*FORCs of a single Py Microstructure*

The above results confirm the interpretation of the 'boomerang' FORC feature put forward by Davies *et al.*[41] and negative FORC features presented by Gilbert *et al.*[26] These results also show that FORC measurements can be performed using microscopy-based techniques and show quantitative agreement with traditional collective magnetometry measurements. Microscopy FORCs present opportunities to perform quantitative measurements on exceedingly small magnetic elements. To demonstrate this, FORC measurements are performed on the aforementioned permalloy microstructure. The rectangular microstructure harbors two flux-closure Landau structures, analogous to a magnetic vortex, as shown in Fig. 3.

The family of MTXM FORCs and FORC distribution for the single microstructure are shown in Figs. 3a and 3b, respectively. The FORC distribution exhibits the two-lobe 'butterfly' feature, similar to magnetometry FORC measurements taken on arrays of vortex-state nanodots.[31] Analogous to the mechanistic reconstruction of the FORC distribution performed above for the boomerang feature, or previously for the wishbone,[26] the reversal sequence of the vortex system, shown in Fig. 3c, can be used to construct the butterfly feature. The top row of the MTXM images, showing the FORC which starts at $\mu_0 H_R$= +40 mT, does not exhibit vortices, but shows minor distortion of the magnetic configuration from the saturated state. Similar to the CoTb sample in saturation, the near-saturated state has $\frac{\partial M}{\partial H} \approx 0$ and contributes very little to the FORC distribution. At $\mu_0 H_R$= 7 mT the FORC branch shows a dramatic 90% reduction in the magnetization at the reversal field (Fig. 3a), signaling vortex nucleation, which is also observed in the MTXM images (2$^{nd}$ row in Fig. 3c). As the field is increased from $H_R$ to positive saturation, the vortex moves transverse to the magnetic field and annihilates from the top and bottom edges of the microstructure. Annihilation from opposite sides indicates that the vortices have opposite chirality,[47] as would be expected to minimize the exchange energy at their meeting point. As this FORC has $\frac{\partial M}{\partial H} > 0$, while every FORC ($\mu_0 H_R > 7$ mT) has $\frac{\partial M}{\partial H} \approx 0$, the second order derivative used to calculate the FORC distribution is non-zero, resulting in a positive feature in the FORC distribution, labeled Feature 1 in Fig. 3b. Next, the FORC branch measured at $\mu_0 H_R$= -40 mT is shown in the middle row of Fig. 3c. The double vortices are now shown displaced from the centerline of the structure, and as the field is increased back towards positive saturation they migrate across the structure, annihilating again at opposite edges of the microstructure. The migration of the vortex across the microstructure is shown to occur reversibly, such that the magnetization is independent of $H_R$, which in-turn implies the FORC distribution is zero.[31] The FORC branch at $\mu_0 H_R$ =-60 mT shows the evolution of the vortex, moving further towards the edge with higher. The evolution of the vortex from $H_R$ to positive saturation shows the vortex again moves across the microstructure. Comparing each field step in $H$ to the prior FORCs ($\mu_0 H_R$ =-40 mT and 7 mT) shows that the images are similar, again consistent with the reversible behavior of the vortex structure.

Next, the FORC branch $\mu_0 H_R$ =-120 mT is considered. This $H_R$ is beyond the saturation field of the microstructure, as indicated by the first image. Increasing $H$ towards positive saturation, $\frac{\partial M}{\partial H} \approx 0$, while the previous branch ($\mu_0 H_R$ =-60 mT) $\frac{\partial M}{\partial H} > 0$, implying the presence of a negative feature in the FORC distribution. While no such feature is observed here, likely due to the coarse field step, it has been observed previously in other vortex systems.[31,47] In this FORC distribution, a new positive feature is present at $\mu_0 H$ =-27 mT, indicating the re-nucleation of the vortices, also observed in the MTXM images. This re-nucleation event results in a positive feature in the FORC distribution, labeled Feature 2. While Feature 1 and Feature 2, representing nominally symmetric events, should be the same magnitude,[31] in this case, Feature 2 is blended with the negative feature due to the course field step, resulting in a visibly weaker feature. As $H$ is increased towards positive saturation, the vortices reversibly migrate across the entire width of the microstructure and annihilate at the opposite edge from their nucleation. This annihilation generates one final feature in the FORC distribution, completing the 'butterfly' FORC feature.

The butterfly-like FORC distribution for the vortex structure is a well-known example of a system with a large reversible component.[31,45,48] Capturing reversible features in the FORC distribution usually requires special analysis beyond the standard derivative since these features first appear on the boundary, $H=H_R$.[40] In one approach the data in this boundary region is treated separately, without the second order derivative.[48] Reversable magnetization is presumably independent of $H_R$, thus this is quite reasonable. Another approach is to extend the dataset, with a so-called constant extension being the most common. The constant extension is achieved by defining $M(H<H_R) \equiv M(H_R)$, which is shown visually in Fig. 3c as the shaded regions to the left of the $H=H_R$ dashed line. The extended data was not measured, but is simply a copy of the $H=H_R$ data. This approach quantitatively reproduces the reversible features.[45] With the extended dataset in-place, the images show a clear dependence in $H_R$ in the shaded regions, but no changes in $H$. Without changes in $H$, each extended region has $\frac{\partial M}{\partial H} = 0$, and the $\partial H_R$ derivative between adjacent zeros is also zero, so no feature appears in the FORC distribution. The extended FORC dataset only manifests a feature at $H=H_R$, at which point there are differences in both $H$ and $H_R$. Thus, the FORC distribution captures the nucleation and annihilation events in the parameter space ($H>H_R$) and the reversible motion of the flux-closure structures at ($H \approx H_R$); the approximate symbol is used because the derivative is performed over the datapoints near the boundary, resulting in a smearing of the feature across the $H=H_R$ boundary.

**Conclusions**

In summary, microscopic magnetization reversal configurations in model systems of CoTb thin film and patterned permalloy element have been studied by first order reversal curve measurements using magnetic transmission X-ray microscopy. The FORC distribution calculated from the images quantitatively agreed with FORCs measured by conventional magnetometry, supporting the microscopic sensitivity of the FORC technique. By comparing the FORC distribution to the MTXM images, the physical origins of the boomerang-shaped feature in the CoTb film were directly confirmed. The ability to perform microscopic FORCs was demonstrated by measuring the reversal of a single Py microstructure with double vortices and calculating the FORC distribution, resulting in the expected butterfly-like FORC feature. Leveraging the element-specific, high spatial resolution characteristics of MTXM, and the quantitative insight from FORC measurements, the combined MTXM-FORC technique could provide new insights into studies of magnetic heterostructures and multi-component complex materials.

**Acknowledgements**

D.A.G was supported by the U.S. Department of Energy, Office of Science, Office of Basic Research CAREER program under Award Number DE-SC0021344. K.L. was supported by the US NSF (DMR-2005108). P.F. was supported by the U.S. Department of Energy, Office of Science, Office of Basic Energy Sciences, Materials Sciences and Engineering Division under Contract No. DE-AC02-05-CH11231 (NEMM program MSMAG). This research utilized resources of the Advanced Light Source, a U.S. DOE Office of Science User Facility under contract no. DE-AC02-05CH11231.

**Figures**

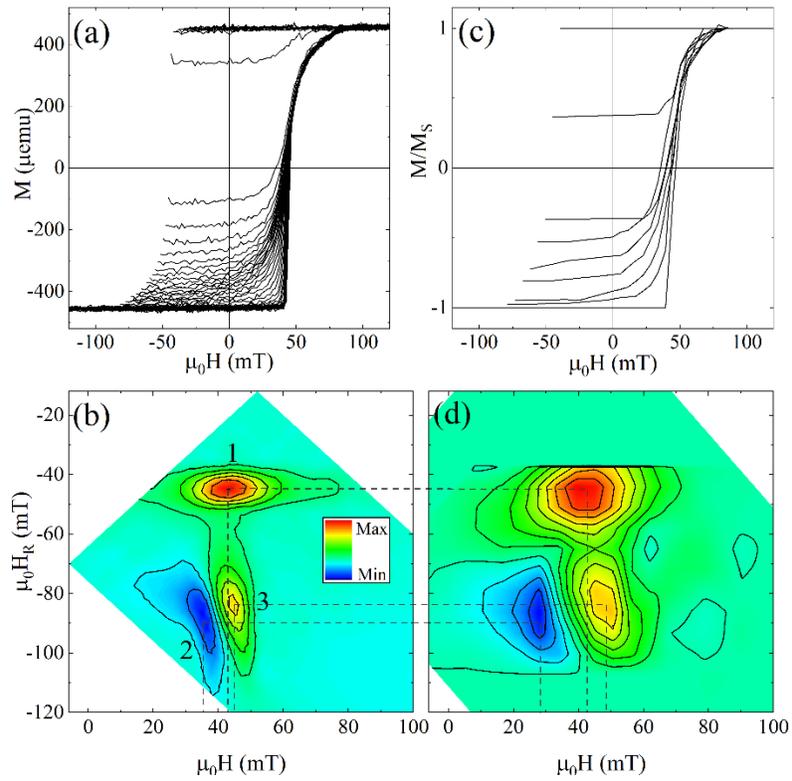

**Figure 1** Family of FORCs and FORC distributions for the CoTb film, as measured with (a, b) AGM and (c, d) MTXM.

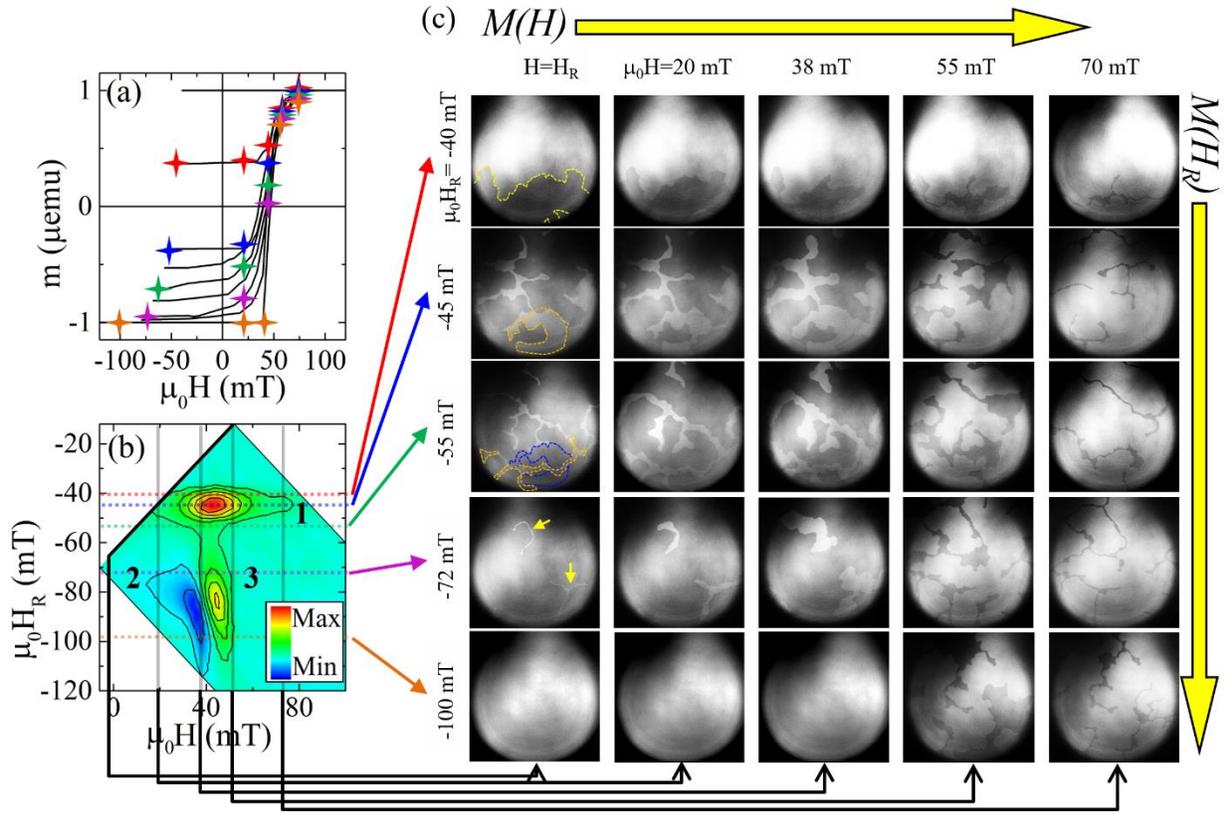

**Figure 2** (a) Family of MTXM FORCs and (b) AGM FORC distributions for CoTb film and (c) corresponding MTXM domain images. The $H_R$ and $H$ values corresponding to each column and row, respectively, are indicated in the FORC diagram. The white (black) contrast indicates positive (negative) out-of-plane oriented domains. The field of view is 10 μm in diameter.

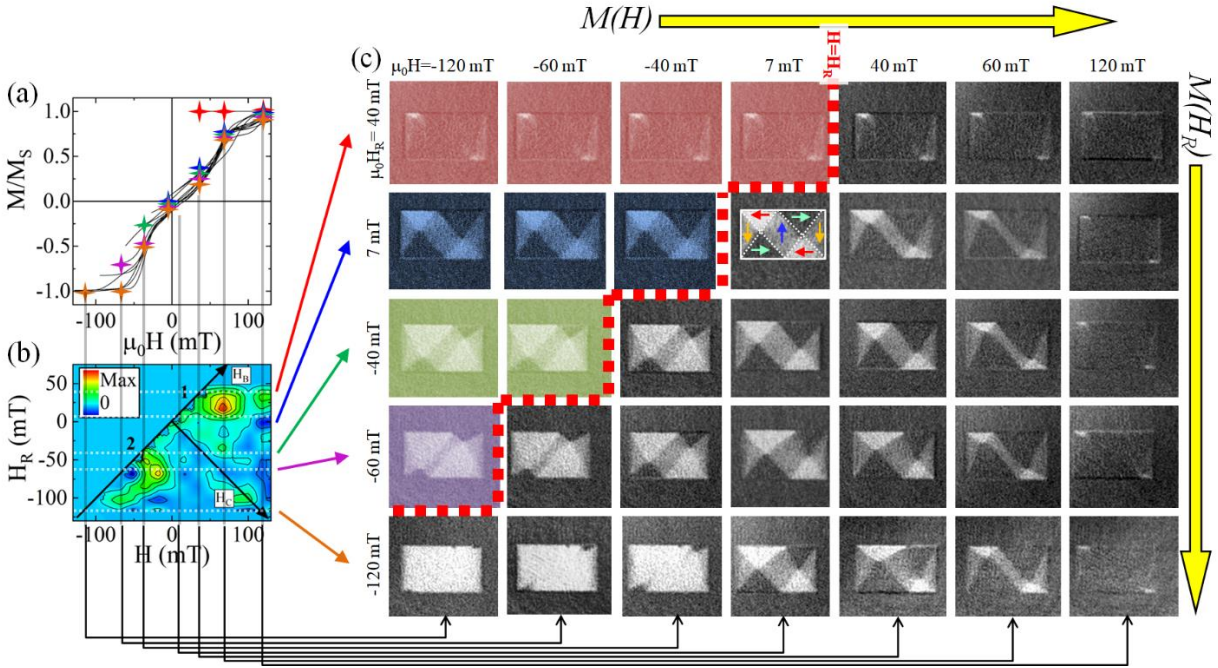

**Figure 3** (a) Family of FORCs and (b) FORC distributions for a 3 μm × 2 μm rectangular Py microstructure as measured with MTXM and (c) corresponding domain images. Arrows in the $\mu_0 H = \mu_0 H_R = 7$ mT indicate the direction of the in-plane magnetization, dotted lines indicate the 90° domain wall between regions of the Landau structure, while the solid white line boarders the nanodot. The $H_R$ and $H$ values corresponding to each column and row, respectively, are indicated in the FORC diagram. The black (white) contrast indicates magnetization parallel (antiparallel) to the positive polarity of the in-plane magnetic field; grey indicates up/down in the image plane.